\documentclass[prb,aps,twocolumn,showpacs]{revtex4}
\usepackage[dvips]{graphicx}
\usepackage{amssymb}
\usepackage{dcolumn}
\usepackage{amsmath}
\usepackage{longtable} 

\newcommand{\MS}{{M_{\mathrm s}}}
\newcommand{\TC}{{T_{\mathrm c}}}
\newcommand{\KB}{{k_{\mathrm B}}}

\begin{document}
\title{Laser induced magnetization switching in films with perpendicular anisotropy: a comparison between measurements and a multi-macrospin model}

\author{C.\ Bunce$^{1}$}
\author{J.\ Wu$^{1}$} \email{jw50@york.ac.uk}
\author{G.\ Ju$^{2}$}
\author{B.\ Lu$^{2}$}
\author{D.\ Hinzke$^{3}$}
\author{N.\ Kazantseva$^{1}$}
\author{U.\ Nowak$^{3}$}
\author{R.\ W.\ Chantrell$^{1}$}

\affiliation{$^1$ Department of Physics, University of York, York YO10
5DD, U.~K.}

\affiliation{$^2$ Seagate Technology, Recording Media Operations, Fremont, CA, U.~S.~A}

\affiliation{$^3$ Fachbereich Physik, Universit\"{a}t Konstanz, 78457 Konstanz,
Germany}

\begin{abstract}
Thermally-assisted ultra-fast magnetization reversal in a DC magnetic field for magnetic multilayer thin films with perpendicular anisotropy has been investigated in the time domain using femtosecond laser heating. The experiment is set-up as an optically pumped stroboscopic Time Resolved Magneto-Optical Kerr Effect magnetometer. It is observed that a modest laser fluence of about 0.3 mJ/cm$^{2}$ induces switching of the magnetization in an applied field much less than the DC coercivity (0.8 T) on the sub-nanosecond time-scale. This switching was thermally-assisted by the energy from the femtosecond pump-pulse. The experimental results are compared with a model based on the Landau Lifschitz Bloch equation. The comparison supports a description of the reversal process as an ultra-fast demagnetization and partial recovery followed by slower thermally activated switching due to the spin system remaining at an elevated temperature after the heating pulse.
\end{abstract}

\pacs{
  75.50.Ss 
  75.40.Mg 
  75.40.gb 
  76.60.es 
}
\maketitle

\section{Introduction}
There is currently considerable interest in ultra-fast laser-induced
magnetization processes. Since the demonstration by Beaurepaire {\em et
al.} \cite{beaurepairePRL96} that the magnetization can respond on the
picosecond timescale to heat pulses produced by femtosecond lasers, a
number of groups \cite{kimelNATURE04,kimelNATURE05,hohlfeldPRB01}
have studied magnetization
processes on this timescale. Experiments generally use pump-probe
processes in which a high energy laser pulse is used to heat the
sample and a low energy probe pulse (split from the main pulse) is
used to monitor the magnetic response using the Magneto-Optical Kerr
Effect (MOKE). Much of this work has investigated the dynamics of
the destruction and recovery of the magnetization, which can occur
on the sub-picosecond timescale, although the recovery can take an
order of magnitude longer due to frustration effects among large
numbers of nucleation sites at which the recovery starts locally
\cite{kazantsevaEPL08}.

The dynamics of reversal during a pulsed
laser experiment in the presence of an applied field has received
less attention.  Hohlfeld {\em et al.} \cite{hohlfeldPRB01} investigated
the magnetization reversal induced by 100 fs laser pulses in a
GdFeCo Magneto-Optical recording medium with perpendicular
anisotropy. They observed an ultra-fast demagnetization of the film
occurring within the first picosecond; followed by a slower
recovery, taking nearly a nanosecond, in the direction of the
applied field as the heat drains from the media layer. They analyzed
their measurements using the Bloch equation and concluded that the
reversal process was governed by the nucleation and growth of
domains in the applied field.

Laser assisted magnetization processes
have considerable potential for future ultra-high density recording
systems. Essentially, the path to higher densities requires
continuous reduction in the grain volume $V$ of the storage
medium, necessitating an increase in the magneto-crystalline
anisotropy energy density $K$ in order to preserve a sufficiently
large value of the parameter $KV/\KB T$ ($\sim 60$) to ensure thermal
stability of written information \cite{weller}. However, the large value of $K$
impacts the writability of the information, and some scheme is
required to overcome this problem. A promising solution is Hybrid or
Heat Assisted Magnetic Recording (HAMR) \cite{rottmayer} in which the medium is heated in
order to lower the anisotropy and thereby allow information to be
written to the medium. Since this is a relatively new field the
magnetization reversal mechanisms are not well understood. Although
the work of Hohlfeld {\em et al.} \cite{hohlfeldPRB01} has demonstrated
thermally activated domain processes in magneto-optical media, perpendicular
recording requires relatively weakly coupled granular media in which
domain processes are not the dominant reversal mechanism.

This paper presents time-domain measurements of the magnetization
reversal process induced by an ultra-fast laser pulse in a
 thin film with perpendicular anisotropy.
The film was especially designed to have a low Curie temperature in
order to simplify the experiments.
We compare the results with a computational model using the
Landau-Lifshitz-Bloch (LLB) equation \cite{garaninPRB97},
which is ideally suited to simulation of
magnetization processes up to and beyond $\TC$ and has been shown
\cite{chubykaloPRB06,atxitiaAPL07} to give an excellent description of the
physics of pulsed laser processes. It is concluded that the
magnetization response consists of a fast demagnetization followed
by a slower response in which the magnetization evolves into the
field direction by a process of thermally activated transitions over
the local energy barriers. Our LLB-micromagnetic model is shown to
give a good description of the physics of the reversal process on
both timescales.

\section{Method}
\begin{figure}[h]
\includegraphics[scale=0.4, angle=0]{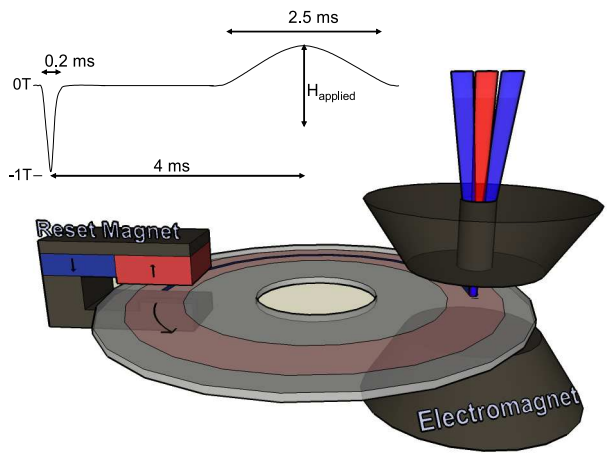}  
\caption{Illustration of spin-stand arrangement used to ensure the materials magnetization is reset between pump pulses.}
\label{f:fig1}
\end{figure}

The experiment is a stroboscopic pump-probe experiment using a
strong laser pulse to initiate a change in the magnetic state of the
sample and a weak probe pulse to observe the resulting magnetization
dynamics via the MOKE. The sample is mounted in a spin-stand which
first moves the magnetic film  through a reset field of magnitude
$\sim 1$T for a duration of  $0.2$ ms which ensures that the
sample is in a remanent state before exposure to each laser pulse.
Fig.~\ref{f:fig1} illustrates the spin-stand arrangement, with
the insert showing the temporal field profile experienced by the
magnetic film. After resetting to saturation the sample then moves
into a perpendicular applied field from an electromagnet (field
range of $\pm$0.52 T). We note that the field applied is always
lower than the static coercivity of the sample as measured at room
temperature by a vibrating sample magnetometer. At the center of the
pole piece a small hole allows optical access to the sample at which
point the sample is exposed to the pump-pulse. The laser pulses
arrive at a rate of 1 kHz but the rotation of the spin-stand (about
7,000 rpm) ensures that freshly saturated film arrives between
pulses.

The stroboscopic experiment uses a Libra laser system  (made by
Coherent) which can produce a 1 kHz stream of 1 mJ, 150 fs pulses of
800 nm radiation. This beam is attenuated, the probe beam is split
off and frequency doubled to 400 nm. The pump beam is routed around
an optical delay line with 17 fs resolution over a 1 ns range and
then focused at normal incidence to a spot approximately 800 $\mu$m
in diameter on the disk surface. The probe beam is polarized and
then brought to a 400 $\mu$m focal spot, centered on the pump beam
with a power level of about 1/20th of the pump. Because the probe is
only half the diameter of the pump it must be noted that there will
be a temperature distribution across the region probed. The probe
approaches the sample surface at near normal incidence; this polar
MOKE geometry yields sensitivity to the out-of-plane component of
magnetization. The reflected probe beam is directed into an optical
bridge detector which uses a Wollaston prism to split the beam into
two orthogonal polarized components which impinge on a two segment
photo-detector. By rotating the detector to the angle where the
output of the two polarization channels (A \& B) are approximately
equal, then the difference between them (A-B) is sensitive to small
changes of the polarization angle of the reflected probe which in
turn is proportional to $m_{z}$. The sum (A+B) is sensitive to
changes in reflectivity $\Delta R$, which is associated with temperature changes and
stress waves due to processes such as lattice expansions.
In order to improve immunity to laser drift
the pump beam was chopped and lock-in amplifiers used to detect the
sum and difference signals. This makes the measurement sensitive to
the difference between the state of the sample without the pump
pulse applied  and the state induced by the pump-pulse. Therefore
the measurements are relative and it is difficult to assign an
absolute scale to the magnetization changes.

\section{Experimental Results}


For characterization of the sample the quasi-static hysteresis loop was measured.
Of particular importance is the
coercivity, which on this timescale is $\sim$ 0.85T, and the
saturation magnetization $\MS$, which is equal to 0.35 $\times 10^6$ A/m. The
coercivity of course is already greater than the field applied
during our pulsed field experiment. However, it is important to note
that the dynamic coercivity on the timescale of the pulsed laser
experiment is even larger. The intrinsic coercivity ($H_0$) and the
thermal stability factor ($KV/\KB T$) were measured by making a series
of time dependent coercivity measurements and fitting to Sharrock's
law \cite{sharrock}. The intrinsic coercivity is related to the
anisotropy field and is expected to give a reasonable estimate of
the coercivity at the nanosecond timescale. The value of $H_0$ was
found to be 1.4T; a factor of almost 3 greater than the maximum
field available from the electromagnet. Separate measurements
determined the anisotropy $K$ to be $4\times 10^6$ J/m$^3$, from
which a grain size of 10nm was estimated.


\begin{figure}[h]
\includegraphics[width=10cm]{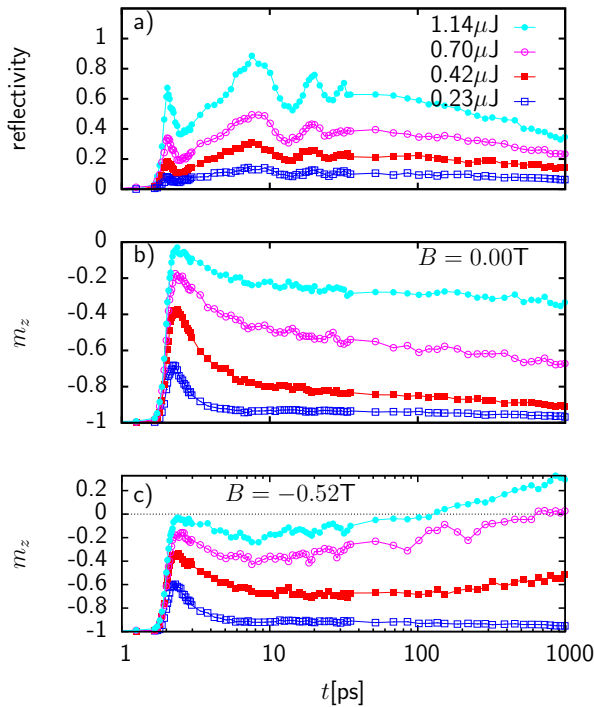}
\caption{Series of measurements in an applied reversing field of 0.52T for different laser pulse energies. a) shows the reflectivity
data, b) the $m_z$ component of magnetization, and c) shows the response
in the presence of a reversing field }
\label{f:fig2}
\end{figure}

Fig.~\ref{f:fig2} shows a set of time-resolved measurements
on  the sample. Fig.~\ref{f:fig2} (a) shows the reflectivity
data, Fig.~\ref{f:fig2} (b) shows the dynamic magnetic response
for zero applied field and Fig.~\ref{f:fig2} (c) shows the
response in the presence of a reversing field of 0.52 T. The laser
pulse energy is varied up to 1.14 $\mu$J per pulse (which corresponds to
a fluence of approximately 0.23 mJ/cm$^{2}$). This value is the
largest than can safely be applied to this sample as energy fluences
above about 0.5 mJ/cm$^{2}$ damage the sample. The sample
reflectivity data shown in Fig.~\ref{f:fig2} (a) is a probe of
the electron/lattice temperature in the system. It indicates that
the same temperature profile is generated each time and that the
temperature scales linearly with pulse energy. The reflectivity
measurement has three distinct features. Within the first 5 ps is a
large peak having a width of 350 fs, which corresponds to the large
rise in electron temperature caused by the arrival of the 150 fs
laser pulse. The electron system then establishes thermal
equilibrium with the lattice which creates the second, rather
broader, peak 20 ps later. In reality the peak temperature the
lattice reaches is much lower than that for the electrons; however,
it seems that in this sample the change in reflectivity is more
sensitive to the lattice temperature than the electron temperature.
The waves seen superimposed onto the lattice temperature peak during
the 3-35 ps time frame are probably stress pulses launched into the
film by the laser heating of the surface, which reflect off the
interface with the glass substrate. The time between peaks is 12 ps,
which, given the film and interlayer thickness of 25 nm, suggests a
propagation speed of 4,100 m/s; consistent with the speed of sound
in a typical metal. The sample appears to cool rather slowly as the
lattice temperature has only fallen to half its peak value after 1
ns. This reflects the fact that in this film there is no heat sink
to absorb the heat. Fig.~\ref{f:fig2} (b) and (c) show the
magnetization dynamics in applied fields of 0 and -0.52 T
respectively. Fig.~\ref{f:fig2} (c) clearly demonstrates heat
assisted reversal due to the pulse. For pulse energies above $\sim
0.5 \mu$J the sample is seen to cool with the magnetization aligning
along the applied field. Recalling that the dynamic coercivity
estimated from magnetic measurements is around 1.46T, this
demonstrates a significant heat-assistance during the sub-nanosecond
reversal process.

We now consider in detail the processes involved in the
magnetization dynamics, which involves three characteristic
timescales as illustrated in Fig.~\ref{f:fig2}. The initial
phase involves a rapid demagnetization of the sample lagging the
change in reflectivity by only 50 fs. The demagnetization takes
about 500 fs, independent of the applied field. This is consistent
with the normal expectation of a rapid demagnetization as previously
demonstrated experimentally \cite{beaurepairePRL96,kimelNATURE05,hohlfeldPRB01} and
theoretically \cite{koopmansJMMM05,kazantsevaEPL08}. The sample then
appears to partially recover its magnetization in the original
direction, on a timescale independent of the applied field. It is
interesting to note that the rate of recovery from the demagnetization
peak is related to the amount of demagnetization achieved - a
pattern consistent with the calculations of Kazantseva {\em et al.}
\cite{kazantsevaEPL08}. Throughout the whole process the only
apparent field dependence occurs in the longer time-scale dynamics
(20ps-1ns). Over this timescale, for the higher laser powers, a
gradual reversal of the magnetization is seen.

\begin{figure}[h]
\includegraphics[bb = 200 310 450 430, width = 8.5cm]{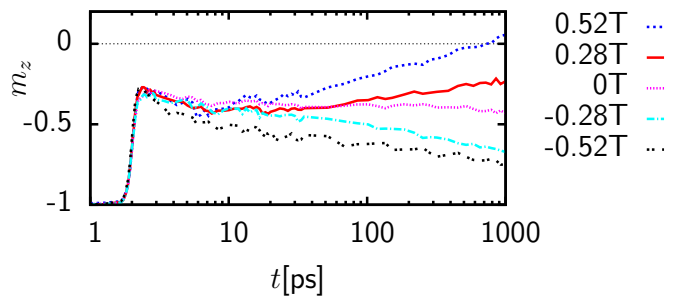}
\caption{Series of measurements of the temporal variation of $m_z$ in various applied reversing fields
 for a constant laser pulse energy.}
\label{f:fig3}
\end{figure}

A similar systematic trend is also exhibited by the temporal
variation of $m_z$ at constant laser power in various applied
fields, as shown in Fig.~\ref{f:fig3}. The initial
disappearance and recovery of the magnetization is similar for all
applied fields, but with increasing positive field the long-term
trend is clearly toward positive magnetization.

The form of the dynamics, involving a trend against the field
direction followed by a slow reversal into the field direction is
apparently somewhat counter-intuitive. However, the form of the
magnetization evolution can be explained by consideration of the
different timescales associated with processes at the atomic and
macrospin length scales. Within each grain or 'macrospin' the
disappearance and recovery of the magnetization will be governed by
the longitudinal relaxation time, which is temperature dependent but
typically of a sub-picosecond timescale, which is consistent with
the experimental data of Fig.~\ref{f:fig2} (b) and (c).
However, after this process the system remains at an elevated
temperature for around 1ns, so there is a possibility of thermally
activated magnetization reversal. This will have a characteristic
timescale determined by the macrospin energy barrier, which is
lowered by the reduction of the anisotropy constant at elevated
temperature, but typically has values much greater than 1ps. On this
basis we would expect to see a fast reduction and recovery of the
magnetization due to atomic processes on the picosecond timescale
with a slower relaxation into the field direction due to thermally
activated reversal of the macrospins. In order to test this
hypothesis in relation to the experimental results we have developed
a computational model of the laser heating process based on the
LLB equation which has been shown \cite{atxitiaAPL07} to give a
good description of magnetization processes over both characteristic
timescales.


\section{Dynamic model of laser heating process} 
\label{s:llb}

We have developed a computational model of laser-induced
magnetization dynamics of a thin film with perpendicular anisotropy.
Consistent with experiments we assume a granular microstructure for
the film. For simplicity we assume, in these initial calculations, a
mono-disperse grain size and anisotropy. Inter-granular magneto-static
interactions are included, but the grains are taken as exchange
decoupled. As mentioned previously the Landau-Lifschitz Gilbert (LLG) equation
cannot be used for models of laser heating since it does not allow longitudinal
 fluctuations of the magnetization\cite{koch,slav}. In the following, we use the
LLB equation \cite{garaninPRB97} for the
thermodynamic simulation of the laser-induced magnetization
switching.  As described in detail in
Ref.~\cite{kazantsevaPRB08,atxitiaAPL07}, the LLB equation has been
shown by comparison with atomistic calculations to give a remarkably
good description of the physics of ultra-fast high temperature
dynamics. The LLB equation can be written as
\begin{eqnarray}
&&\mathbf{\dot{m}}_i=-\tilde{\gamma} \lbrack \mathbf{m}_i\times  \mathbf{H}^{i}_{%
\mathrm{eff}} ]+\frac{\tilde{\gamma} \alpha _{||}}{m_i^{2}}\left(%
\mathbf{m}_i\cdot ( \mathbf{H}^{i}_{\mathrm{eff}}+\boldsymbol{\zeta }^{i}_{||})\right) %
\mathbf{m}_i \nonumber \\
&&\qquad {}-\frac{\tilde{\gamma} \alpha
_{\perp}}{m_i^{2}} \left[\mathbf{m}_i\times \left \lbrack
\mathbf{m}_i\times \left( \mathbf{H}^{i}_{\mathrm{eff}}+\boldsymbol{\zeta}^{i}_{\perp}\right) %
\right] \right]. \label{e:llb}
\end{eqnarray}

Note, that besides the usual precession and relaxation terms in the
LLG equation  (see Ref.~\cite{nowakBOOK07}
for more details), the LLB equation contains an additional term
which controls the longitudinal relaxation. Here, ${\bf m}_i$ is a
spin polarization which is not assumed to be of constant length and
even its equilibrium value $m_{\rm e}(T)$ is temperature dependent.
${\alpha_{\parallel}}$ and $\alpha_{\perp}$ are dimensionless
longitudinal and transverse damping parameters.

The LLB equation is valid for finite temperatures and even above
$\TC$ though  the damping parameters and effective fields are
different below and above $\TC$. For $T \leq \TC$ the damping
parameters are
\begin{equation}
  {\alpha_{\parallel}} = \lambda \frac{2  T}{3 \TC} \quad
   \alpha_{\perp} = \lambda (1 - \frac{T}{3 \TC})
\end{equation}
and for $T \geq \TC$ the damping parameters are equal,
\begin{equation}
  \alpha_{\perp} = \alpha_{\parallel} = \frac{2 \lambda T}{3 \TC}.
\end{equation}
In these equations $\lambda$ is a microscopic parameter which
characterizes the coupling of the individual, atomistic spins with
the heat bath.

Thermal fluctuations \cite{garaninPRB04} are included as an
additional noise term ${\boldsymbol \zeta}^i_{l}(t)$ with $l = \perp, \parallel$, $\langle {\boldsymbol \zeta}^i_{l}(t) \rangle = 0$ and
\begin{equation}
\label{stocfield}
\langle {\zeta}_{l}^{i,\nu}(0){\zeta}_{l}^{j,\eta}(t) \rangle = { 2k_B T \over
\tilde{\gamma} \alpha_{l} M_{\mathrm s}^0 \Delta^3} \delta_{\nu \eta} \delta_{ij} \delta(t),
\end{equation}
where $i,j$ denotes lattice sites and $\nu,\eta$ the Cartesian components. Here, $\Delta^3$ is
the volume of the grains and $M_{\mathrm s}^0$ is the value of the spontaneous
magnetization at zero temperature.

The effective fields are $ \mathbf{H}_{\mathrm{eff}}^i = -
\frac{1}{M_{\mathrm s}^0} \frac{\delta f}{\delta \bf{m}_i}$, with
$f$ the free energy density. The total local field is given by \cite{garaninPRB97}
\begin{equation}
  \mathbf{H}_{\mathrm{eff}}^i = \mathbf{H}+\mathbf{H}_{A}^i +\mathbf{H}_{\rm dipol}^i + \left\{
  \begin{array}{cc}
    \; \; \; \; \; \; \; \;
     \frac{1}{2\tilde{\chi}_{\Vert }}\left(
     1-\frac{m^{2}_i}{m_{\rm e}^{2}}\right) \mathbf{m}_i & T \leq \TC \\
    -\frac{1}{\tilde{\chi}_{\Vert }}\left(1 +  \frac{3\TC m^{2}_i}{5(T - \TC)}\right)
    \mathbf{m}_i & T \geq T_{\mathrm c}
  \end{array}
  \right. \mathbf{.}
  \label{Heffm}
\end{equation}
with the anisotropy field
\begin{equation}
  \mathbf{H}_{A}^i  =  -\frac{\left( m_{x}^i\mathbf{e}_{x}+m_{y}^i\mathbf{e}_{y}\right)}{\tilde{\chi}_{\perp}}
\end{equation}
which makes the $z$ axis the easy axis of the model, and the dipolar field
\begin{equation}
 \mathbf{H}_{\rm dipol}^i =   \frac{M_{\mathrm s}^0 \Delta^3 \mu_0}{4 \pi}\sum\limits_{i<j}
           \textstyle \frac{{3 ({\bf m}_i
    \cdot {\bf e}_{ij})({\bf e}_{ij} \cdot {\bf m}_j) -
    \bf m}_i \cdot {\bf m}_j} {r_{ij}^3}
\end{equation}

\begin{figure}
\includegraphics[bb = 180 270 460 440, width = 8cm]{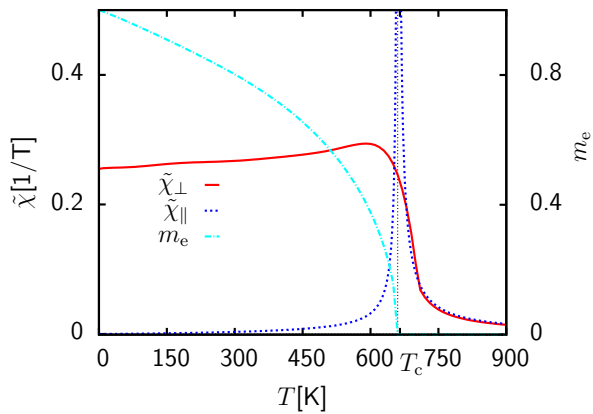}
\caption{Spontaneous equilibrium magnetization and parallel and transverse susceptibility vs.
temperature.}
\label{f:fig4}
\end{figure}

Note, that the susceptibilities $\tilde{\chi}_{l}$ are defined by
$\tilde{\chi}_{l} = \partial m_l / \partial H_l$. At lower
temperatures the perpendicular susceptibility
 $\tilde{\chi}_{\perp}$ is related to the anisotropy $K$ via
 $\tilde{\chi}_{\perp} = M_{\mathrm s}^0 m_{\rm e}^2/(2K)$. \cite{garaninPRB97}.
One problem for the application of the LLB equation is that one
has to know the functions for the spontaneous equilibrium
 magnetization $m_{\rm e}(T)$, the perpendicular ($\tilde{\chi}_{\perp}(T)$)
and parallel ($\tilde{\chi}_{\parallel}(T)$) susceptibilities. Here,
we use the functions for FePt gained from a Langevin dynamics simulation
of an atomistic spin model as described in \cite{kazantsevaPRB08}.
In the following, we assume that these functions reflect the correct
temperature behavior. We normalize the perpendicular susceptibility such that
its value at 300K is consistent with experimentally determined values
for CoPt ($K = 3.94 \cdot 10^5$ J/m$^3$) determined by the rotation method \cite{Lu}.
The input functions are shown in Fig. \ref{f:fig4}.

The LLB equation is solved numerically by using Langevin dynamics
simulations as described in \cite{nowakBOOK07}. For our simulations
we chose a disc  of $32 \times 32 \times 1$ cells, with a
grain size $\Delta$ of 10nm.

\section{Comparison between Multi-Macro Spin Model and Experiment}
In order to make a comparison with the experimental data it is
necessary to have an approximation to the temporal variation of the
temperature pulse caused by the heating. Because of the complex behavior of the
reflectivity as discussed by Djordjevic et al. \cite{djordjevic} it is not feasible to use
this property to determine the temperature profile in this work. Instead we make the
simplifying assumption that, at low laser powers, the magnetization
closely follows the electron temperature during the heating and
initial recovery phase; an assumption essentially borne out by
calculations in Ref. \cite{kazantsevaEPL08} using an atomistic
model, where fast demagnetization and recovery were found for low
peak electron temperatures. Following Ref.
\cite{kazantsevaEPL08}, we assume that the photon energy is
transfered to electrons and that the magnetization is directly
coupled to the electron temperature $T_{\mathrm e}$ within a two-temperature
model \cite{zhangBOOK02}, expressed as the following coupled differential equations for
the electron and phonon temperatures, $T_{\mathrm e},T_{\mathrm l}$ respectively:
\begin{eqnarray}
C_{\mathrm e} {{\mathrm d} T_{\mathrm e} \over {\mathrm d}t} &=& -G_{\mathrm el} (T_{\mathrm e} - T_{\mathrm l}) + P(t) \nonumber \\
C_{\mathrm l} {{\mathrm d} T_{\mathrm l} \over {\mathrm d}t} &=& G_{\mathrm el} (T_{\mathrm e} - T_{\mathrm l}),
 \label{e:temp}
\end{eqnarray}
where $C_{\mathrm e}, C_{\mathrm l}$ are electron and lattice specific heat constant,
$G_{\mathrm el}$ is a coupling constant and $P(t)$ is the laser fluence.
Eqs.~\ref{e:temp} can easily be solved numerically to generate the
time variation of $T_{\mathrm e},T_{\mathrm l}$. We determine the parameters of the
two-temperature model by fitting to the form of the initial
magnetization decay and recovery, assuming a Gaussian laser profile.

\begin{figure}
\includegraphics[bb=180 270 460 457, width = 8cm]{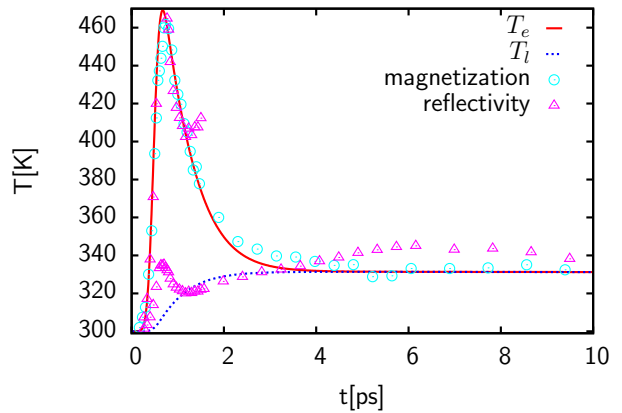}
\caption{Graph showing the electron temperature ($T_{\mathrm e}$) and the
lattice temperature ($T_{\mathrm l}$)  simulated by the two-temperature
model. Scaled reflectivity and magnetization data are also included
for comparison.}
\label{f:fig6}
\end{figure}

Fig. \ref{f:fig6} shows a comparison of the two-temperature
model with the data from the reflectivity and low laser power
magnetization measurements in order to estimate physically
reasonable parameters for the model. The time-scale of the peak in
electron temperature $T_{\mathrm e}$ matches the demagnetization peak in the
magnetization data and the initial peak in the reflectivity data.
The lattice temperature $T_{\mathrm l}$ reaches its peak value on the same
time-scale as the second rise in the reflectivity data. For use in
the computational model an interpolation function was fitted to the
$T_{\mathrm e}$ predicted above. Different laser fluences were simulated by
scaling the results to the peak electron temperature ($T_{\rm e}^{\rm p}$), which becomes
a main parameter in the comparison with experiment.

\begin{figure}
\includegraphics[bb = 160 245 500 500,width = 9cm]{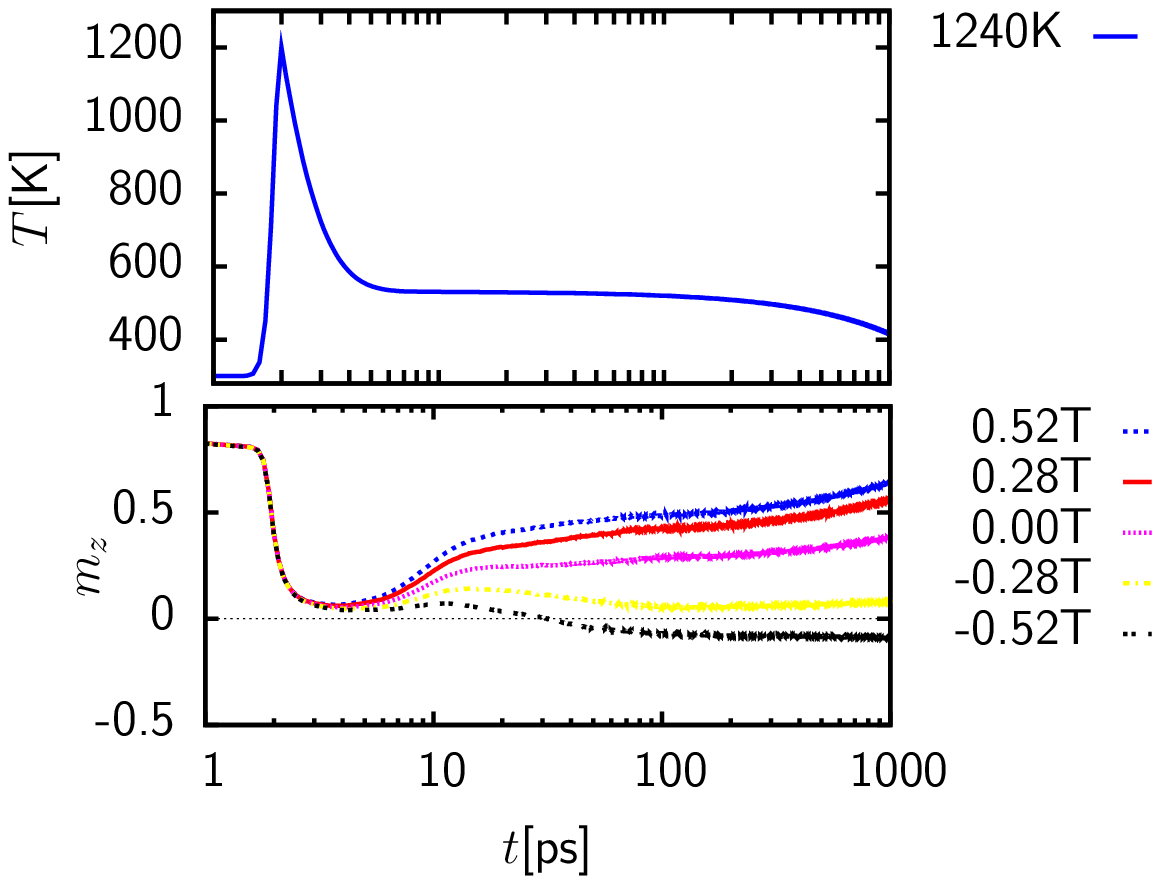}
\caption{Calculated time dependence of the $z$-component of the
magnetization for different applied magnetic fields.
         $\frac{KV}{\KB T} = 76$ with K = $3.2 \cdot 10^5$ J/m$^3$ and $V$ = (10 nm)$^3$ at room temperature. ($T_{\rm e}^{\rm p}$ = 1240K, $\lambda = 0.1$)}
\label{f:fig7}
\end{figure}

We first describe calculations with the assumption of a spatially
uniform temperature profile. The model parameters used correspond to
a material with a $\TC$ of 660K, $\MS$ = 1.75 $\times10^{5}$  A/m,
 an out-of-plane anisotropy $K$ = 3.2 $\times 10^{5}$ J/m$^3$
and a damping constant $\lambda$ of 0.1.
The material is broken up into $32 \times 32$ cells of size 10 nm which are exchange
decoupled in order to model the granular structure of a recording
medium. The intention is to outline the effects of the major
parameters in the model, namely the applied field, peak electron
temperature and the coupling parameter.
Fig. \ref{f:fig7} shows the temporal response of the magnetization
to a laser pulse giving rise to a peak electron temperature
of $1240K$ assuming a coupling parameter of $\lambda=0.1$. It can be seen that
the simulation gives a reasonable qualitative description of the time evolution of the
magnetization following a laser pulse. Specifically we note the
initial fast demagnetization and recovery. In the case of a positive
field the magnetization recovers to the equilibrium value in the
positive sense. In a reversing field the initial recovery is
followed by a slow reversal of the magnetization toward the field
direction which, in the model, can be unambiguously attributed to
thermally activated transitions over the particle energy barriers,
supporting the earlier interpretation of the experimental data.

\begin{figure}
\includegraphics[bb = 200 300 440 430,width = 8cm]{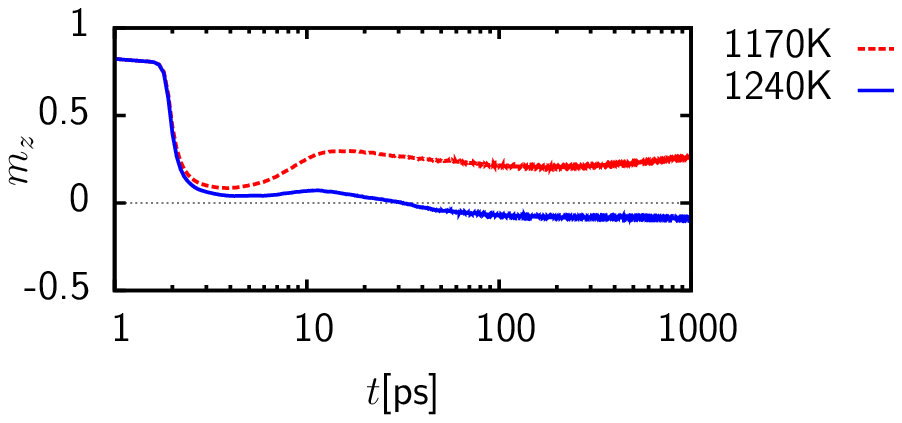}
\caption{Calculated time dependence of the $z$-component of the
magnetization for two different peak temperatures.
$\frac{KV}{\KB T} = 76$ with K = $3.2 \cdot 10^5$ J/m$^3$ and $V$ = (10 nm)$^3$ at room temperature. ($\lambda = 0.1$,$B_z = -0.52$T)}
\label{f:fig8}
\end{figure}

Fig. \ref{f:fig8} explores the effect of increasing the peak
electron temperature. As might be expected, the effect of this
parameter is to lead to a more complete demagnetization during the
laser pulse. After the high temperature pulse the system remains
demagnetized due to the high temperature remaining after the pulse.

\begin{figure}
\includegraphics[bb = 200 300 440 430,width = 8cm]{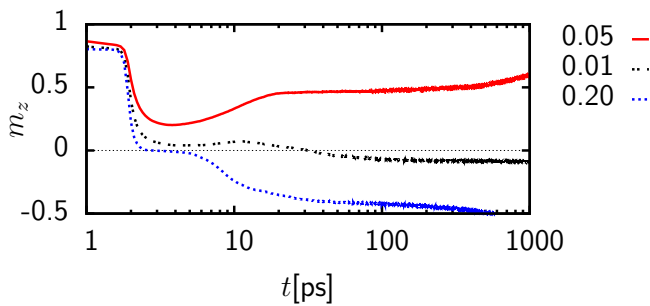}
\caption{Calculated time dependence of the $z$-component of the
magnetization for different values of $\lambda$.
         $\frac{KV}{\KB T} = 76$ with K = $3.2 \cdot 10^5$ J/m$^3$ and $V$ = (10 nm)$^3$ at room temperature ($T_{\rm e}^{\rm{p}} $ = 1240 K, $B_z = -0.52$T)}
\label{f:fig9}
\end{figure}

Finally, we consider the effect of the coupling parameter $\lambda$.
Calculations for 3 different values of $\lambda$ are shown in Fig.
\ref{f:fig9}. Here, it can be seen that the effect of increasing
$\lambda$ is to achieve a more rapid demagnetization. This is
consistent with previous calculations \cite{kazantsevaEPL08}, where
the effect is interpreted in terms of the more efficient transfer of
energy into the spin system at large $\lambda$ leading to a more
complete demagnetization for a given laser pulse width. Clearly the
increased demagnetization caused by the stronger coupling to the
conduction electron system results in an increased heat assistance
of the magnetization reversal.

In addition to the material parameters, the peak electron
temperature and the coupling constant are seen to be important in
the heat assisted reversal process. Although the model reproduces
the essential physics of the reversal process, particularly the
different behavior on the timescales of longitudinal relaxation
(fast demagnetization) and transverse relaxation (super-paramagnetic
fluctuations) qualitative agreement only is obtained with
experimental data. Specifically, the value of $KV/\KB T=76$ (at room temperature) used in the
simulations is rather small in comparison with the experimentally
determined values. In order to obtain closed agreement with
experiment it is necessary to include an important experimental
factor; specifically the fact that the probe pulse area is
comparable to that of the pump. This introduces a significant
temperature gradient within the probe area, which must be taken into
account in the calculations.

Here, we model this effect using a Gaussian temperature profile for
the laser spot. The diameter of the spot used in the model
calculations is much smaller than in the experiments ($800
\mu m$). This would be an unrealistic approximation if the
magnetization reversal involved domain wall processes; however,
since the grains are essentially decoupled in the experimental films
this simplified model is able to gives some insight into the effects
of the temperature profile. Since both pump and probe beams have a
Gaussian temperature variation, we take the temperature variation to
be of the form $T(r)\propto \exp(-r^2/r^2_{\mathrm{pump}})$, with $r_{\mathrm{pump}}$
the radius of the pump beam. The MOKE signal is assumed to be
proportional to the product of the magnetization with a sensitivity
function. The sensitivity function is proportional to the area of material generating the MOKE signal at a particular radius and the light intensity at that radius, which has a
Gaussian weighting, i.e. $\propto m_z r\exp(-r^2/r^2_{\mathrm{probe}})dr$,
with $r_{\mathrm{probe}}$ the radius of the probe. A numerical integration
over the probe area was carried out to determine the calculated MOKE
signal from the film.

\begin{figure}
\includegraphics[bb = 200 300 440 430,width = 8cm]{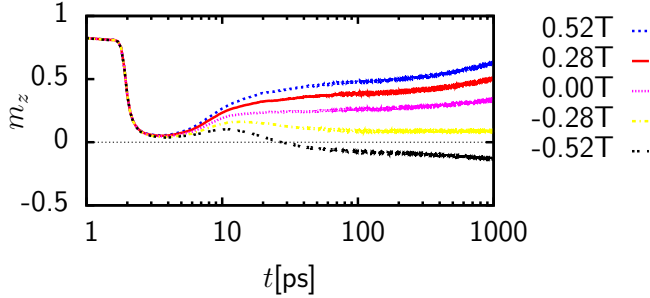}
\caption{$z$-component vs. time for different applied magnetic fields $B_z$.
         $\frac{KV}{\KB T} = 95$ with K = $3.94 \cdot 10^5$ J/m$^3$ and $V$ = (10 nm)$^3$ at room temperature ($T_{\rm e}^{\rm{p}} $ = 1480K, $\lambda = 0.1$)}
\label{f:fig10}
\end{figure}



The introduction of a temperature profile results in a qualitative
agreement with experiment using parameters close to the measured
values. Fig. \ref{f:fig10} shows calculations of the time
dependence of the magnetization following a laser pulse with a peak
temperature of $1480$K for different values of the applied magnetic field.
The parameters used were $M_{\mathrm s}^{0}=0.4\times 10^{6}$ A/m, giving a room temperature
 value close to the measured
(VSM) values of $0.32\times10^{6}$ A/m. The anisotropy constant
$K^0$ at zero temperature was taken as $3.94\times 10^{5}$J/m$^{3}$, which
gives a value of $KV/\KB T=95$ at room temperature, in good agreement
with the experimentally determined value of 96.
It is interesting to note that if a larger value of $M_{\mathrm s}^{0}$ is used then system appears to lock
into a domain-state during reversal, which suggests that even at
elevated temperatures the inter-granular magneto-static interactions
can play an important role. The exchange coupling can be added, but
values as big as 10\% do not change the results significantly. As noted by Kazantseva {\em et al.} \cite{kazantsevaEPL08}, changes in the value of the damping constant affect the amount of energy the
magnetic system absorbs from the initial heat pulse and so how much
of a demagnetization is achieved for a given peak electron temperature. In addition there is
a slight broadening in the demagnetization peak.

\begin{figure}
\vspace{2cm}
\includegraphics[bb = 200 70 440 600,width = 5cm]{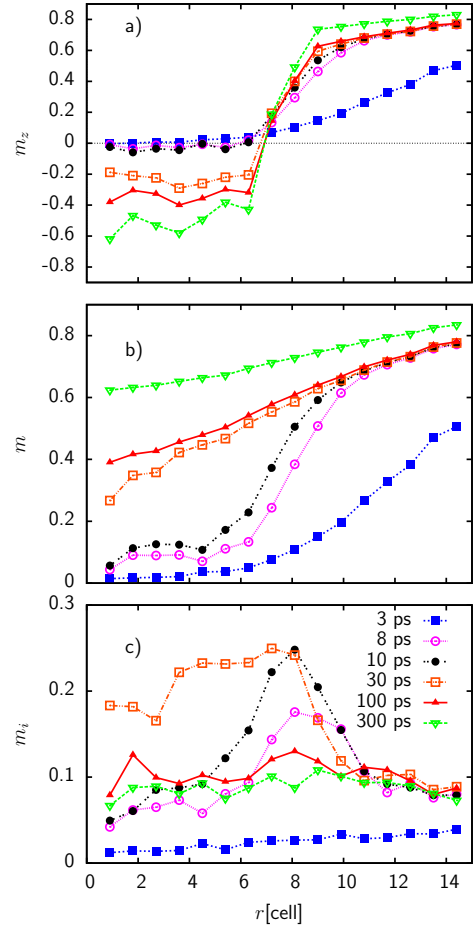}
\caption{Radial magnetization in a 0.52T field as a function of time after the laser pulse. (a) shows the perpendicular component $m_z$, (b) the magnitude $m $, and
(c) the magnitude of the in-plane component $m_i$}
\label{f:fig12}
\end{figure}

Fig.~\ref{f:fig12} shows a more detailed analysis of the LLB
simulations,  indicating that the magnetization dynamics are an
ultra-fast demagnetization and recovery caused by the electron
temperature peak, after which the elevated temperature of the
lattice causes a gradual switching of the individual grains of
material. Fig.~\ref{f:fig12} shows the temporal evolution of the radial magnetization defined such that
$m_\alpha(r) dr$ is the spatially averaged magnetization over the annulus $r \rightarrow r+dr$; here $\alpha$ is the z-component of the magnetization $m_z$ (Fig.~\ref{f:fig12}(a)), the total magnetization (Fig.~\ref{f:fig12}(b)), and the in-plane magnetization, defined as the spatial average of $(m_x^2+m_y^2)^{1/2}$ (Fig.~\ref{f:fig12}(c)).

We consider the behavior of two distinct regions; the central region for radius $<$ 7 cell radii, where the temperature exceeds $\TC$ during the pulse leading to complete demagnetization, and the outer region which doesn't exceed $\TC$. The variation of the total magnetization (Fig.~\ref{f:fig12}(b)) is consistent with previous calculations \cite{kazantsevaEPL08}. Specifically, the rate of recovery of the magnetization depends upon the magnetic state. Within the central region the material is completely disordered and the recovery of the magnetization is relatively slow due to the need for the magnetization to recover from highly disordered states. In the outer region the magnetization retains some memory of the initial state, which results in a rapid recovery \cite{kazantsevaEPL08}.

Of most importance in terms of heat assisted reversal is the behavior of the central region. Heat assistance of the reversal is demonstrated clearly in Fig. \ref{f:fig12}(a), which shows reversal of the central (heated) region while the magnetization in the outer region is not switched. The nature of the reversal in the central region is further investigated using the radially resolved in-plane component of the magnetization, which is shown in Fig. \ref{f:fig12}(c). It is interesting to note that a large in-plane component develops on the timescale of $10 \rightarrow 30$ ps. This results from the relatively random recovery of the direction of the magnetization after cooling though $\TC$. This contributes to the magnetization reversal in two ways. Firstly, some of the grains take on a negative sense of the magnetization on recovery. Others will recover in a positive sense but at an angle greater than the energy maximum as the anisotropy increases; these grains are most likely to switch into the negative direction as the temperature decreases. At longer timescale, and consequently lower temperatures, the in-plane component reduces as the magnetization of each grain begins to lie preferentially along the  easy anisotropy axis. However, the in-plane component does not completely disappear, probably reflecting the Boltzmann distribution of the magnetization direction within each grain. In addition to these mechanisms it is also likely that there will be thermally activated reversal over the energy barriers at the elevated temperatures. Along with the increase in spontaneous magnetization as the film cools this would contribute to the gradual increase in $m_z$ over timescales of 1 ns.

\section{Conclusions}
We have presented laser pump-probe measurements which show a clear heat assistance from the laser
pulse for switching the magnetization state. This is demonstrated by the ability to switch in an externally applied
field with a magnitude lower than the intrinsic coercivity. The experiments show a rapid demagnetization and recovery followed by a slow evolution of the magnetization into the field direction. This is consistent with the existence of two characteristic relaxation times; the longitudinal and transverse relaxation times. The former is atomic-scale processes and is typically of the sub-picosecond order, whereas the transverse relaxation time reflects transitions over the energy barrier and can be orders of magnitude longer. In order to investigate the reversal mechanism we have developed a micromagnetic model based on the LLB equation which naturally includes both timescales.  The  LLB model calculations are in good quantitative agreement with the experimental data, as long as the the temperature gradient across the probe pulse is included.  It appears
that on the short time-scale (2 ps) there is a rapid demagnetization of $m_{z}$
 due to an associated loss of $\MS$ via the longitudinal relaxation. There is a partial recovery of
$m_{z}$ in the original direction as $\MS$ starts to recover. However, the switching is assisted by the recovery of the magnetization of individual grains in random directions as the system cools through $\TC$. On the longer time-scale the reversal of
$m_{z}$ in the applied field may also be assisted by thermally activated switching
caused by the elevated lattice temperature. The elevated temperature has the effect of lowering the anisotropy energy barriers (due to the reduced values of the Magnetocrystalline anisotropy energy) and also provides the thermal energy to induce the transitions. The complex behaviour requires a model including both the longitudinal and transverse relaxation times, which is included here using the LLB equation. Our LLB based calculations encapsulate the physics of the heat-assisted reversal process and suggest the LLB equation as a physically realistic model for Heat Assisted Magnetic Recording.

\section{Acknowledgments}
The authors would like to acknowledge the CLF (Central Laser
Facility)  who loaned the laser system used in the project and Dr. S. Lepadatu for his assistance in developing the experiment.



\end{document}